\documentclass{aastex}
\usepackage{epsfig}
\usepackage{emulateapj5}

\newcommand{\Os}{\ion{O}{6}}
\newcommand{\Ct}{\ion{C}{3}}
\newcommand{\hh}{H$_{2}$}
\newcommand{\kms}{km~s$^{-1}$}
\newcommand{\N}{\ion{N}{1}}

\shorttitle{\emph{FUSE} and STIS observations of AB Aur}
\shortauthors{Roberge et al.}

\begin{document}

\title{\emph{FUSE} and \emph{HST} STIS Observations
of Hot and Cold Gas in the AB Aurigae System}

\author{A.~Roberge\altaffilmark{1}, A.~Lecavelier des Etangs\altaffilmark{2},
C.~A.~Grady\altaffilmark{3}, A.~Vidal-Madjar\altaffilmark{2}, 
J.-C. Bouret\altaffilmark{3},
P.~D.~Feldman\altaffilmark{1}, M.~Deleuil\altaffilmark{4}, 
M.~Andre\altaffilmark{1}, A.~Boggess\altaffilmark{5}, 
F.~C.~Bruhweiler\altaffilmark{5}, R.~Ferlet\altaffilmark{2}, 
and B.~Woodgate\altaffilmark{3}} 
\email{akir@pha.jhu.edu, lecaveli@iap.fr, cgrady@echelle.gsfc.nasa.gov, 
mandre@pha.jhu.edu, magali.deleuil@astrsp-mrs.fr, pdf@pha.jhu.edu,
ferlet@iap.fr, alfred@iap.fr, woodgate@stis.gsfc.nasa.gov}

\altaffiltext{1}{Department of Physics and Astronomy, Johns Hopkins University,
Baltimore, Maryland 21218}

\altaffiltext{2}{Institut d'Astrophysique de Paris, CNRS, 98bis Bd Arago, 
F-75014 Paris, France}

\altaffiltext{3}{NASA/Goddard Space Flight Center, Code 685, 
Greenbelt, Maryland 20771}

\altaffiltext{4}{Laboratoire d'Astronomie de Marseille, BP 8, F-13376 
Marseille Cedex 12, France}

\altaffiltext{5}{Institute for Astrophysics \& Computational Sciences,
Department of Physics, The Catholic University of America, Washington, 
DC 20064}

\begin{abstract}

We present the first observations of a Herbig Ae star with a 
circumstellar disk by the \emph{Far Ultraviolet Spectroscopic Explorer} 
(\emph{FUSE}), as well as a simultaneous observation of the star obtained 
with the \emph{Hubble Space Telescope} Space Telescope Imaging
Spectrograph (STIS). 
The spectra of AB Aurigae show emission and absorption features 
arising from gasses that have a wide range in temperature, from hot \Os\ 
emission to cold \hh\ and CO absorption. Emissions from the 
highly ionized species \Os\ and \Ct\ present in the \emph{FUSE} spectrum
are redshifted, while absorption features arising from 
low-ionization species like \ion{O}{1}, \N, and \ion{Si}{2} are 
blueshifted and show characteristic stellar wind line-profiles. 
We find the total column density of \hh\ toward AB Aur from the \emph{FUSE}
spectrum, $N(\rm{H}_2) = (6.8 \pm 0.5) \times 10^{19}$ cm$^{-2}$. 
The gas kinetic temperature of the \hh\ derived from the ratio 
$N(J=1)/N(J=0)$ is $65 \pm 4$ K. The column density of the CO observed 
in the STIS spectrum is $N(\rm{CO}) = (7.1 \pm 0.5) \times 10^{13}$ 
cm$^{-2}$, giving a CO$/$\hh\ ratio of $(1.04 \pm 0.11) \times 10^{-6}$. 
We also use the STIS spectrum to find the column density of \ion{H}{1},
permitting us to calculate the total column density of hydrogen atoms, 
the fractional abundance of \hh, and the gas-to-dust ratio.

\end{abstract}


\section{Introduction}

AB Aurigae (HD 31293) is one of the brightest and best studied Herbig 
Ae stars, which 
are considered to be pre-main-sequence stars of intermediate mass 
(about 2 -- 10 M$_{\odot}$) \citep{wat98}. The star (spectral type A0Ve+sh) 
is about 2 Myr old and is located at a distance of 144$^{+23}_{-17}$ pc 
from the Sun \citep{van98}. Like most Herbig Ae/Be stars, AB Aur is 
surrounded by circumstellar gas and dust, some of which is distributed 
in a disk about the star. Standard 
stellar theory predicts that Herbig Ae/Be stars should not have convective 
layers, and therefore, should not have chromospheres, coronae, or strong
stellar winds. However, AB Aur and most other Herbig Ae/Be stars do show 
chromospheric and wind features \citep{boh93}. The \emph{Far Ultraviolet 
Spectroscopic Explorer} (\emph{FUSE}) provides high-resolution 
access to a portion of the AB Aur spectrum previously observed only at 
low spectral resolution with the 
\emph{Hopkins Ultraviolet Telescope}. In this Letter, we present  
\emph{FUSE} observations of AB Aur, including the first observation of 
hot \Os\ gas ($T \approx 3 \times 10^{5}$ K) and cold molecular hydrogen. 
The \Os\ emission should provide an important constraint on any models 
of the stellar activity seen in the AB Aur system.
We also present a simultaneous observation of AB Aur with the 
\emph{Hubble Space Telescope} (\emph{HST}) Space Telescope Imaging 
Spectrograph (STIS); this observation strongly complements the 
\emph{FUSE} observations by providing column densities of \ion{H}{1} 
and CO. 

\section{\emph{FUSE} observations and data reduction}

AB Aurigae was observed with \emph{FUSE} on 2000 February 27 and 28, 
for a total exposure time of 13.7 ks. 
The data were obtained using the low-resolution aperture (LWRS), 
and alignment of the four channels (LiF 1, LiF 2, SiC 1, and SiC 2) was 
maintained throughout the whole observation (see \citet{sah00} for 
a more detailed explanation). The data were 
flux calibrated using \emph{calfuse 1.6.8} pipeline processing
software, but more recent wavelength solutions (2000 June 19) were 
used for our analysis. We obtained a S/N of about 5 per 12-pixel 
resolution element in the LiF 1a segment (near 1060 \AA), and about 7 
in the LiF 1b segment (near 1140 \AA). 

The wavelength solutions used provide good relative calibration 
across the LiF channels. However, they have significant zero-point offsets. 
In order to establish the absolute wavelength calibration of our spectra, 
we made use of the simultaneous STIS observation of AB Aur described 
in \S 5. Assuming that our \hh\ gas is located in the same region and 
is at the same velocity as the carbon monoxide gas observed in the STIS 
spectrum, we set $v_{\rm{H}_{2}} = v_{\rm{CO}} = 10 \pm 3$ \kms, which is 
close to the heliocentric radial velocity of the star 
($v_{\rm{rad}} = 8$ \kms).

\section{Analysis of \emph{FUSE} spectrum}

The spectrum in Figure~1a shows signal from AB Aur down to 
\Ct\ at 977 \AA, and is rich in emission and absorption 
features, many of which we have not yet been able to identify. The 
stellar photosphere flux drops off sharply at around 1270 \AA\ (Figure~1b),
so virtually all the observed flux in the \emph{FUSE} bandpass is due
to a forest of emission lines and possibly some far-UV excess continuum 
of uncertain origin.   

\subsection{Emission from highly-ionized species \Os\ and \Ct}

The \Os\ emission doublet in the LiF 1a segment spectrum is shown in 
Figure~2a. The $\lambda$1038 line is suppressed by \hh\ absorption; 
two narrow \hh\ lines are also superimposed on the $\lambda$1032 line. These 
two narrow lines were masked out of the data, and a Gaussian model was 
least-squares fitted to the spectrum between 1029 \AA\ and 1035 \AA.
A model for the \Os\ doublet was created using 
this Gaussian fit, assuming the optically thin line ratio 2:1 to
model the $\lambda$1038 line. This \Os\ model was multiplied by 
the normalized \hh\ absorption model discussed in \S 3.2; the
result is overplotted on the data in Figure~2a, showing how the strong \hh\ 
absorption lines near 1037 \AA\ and 1039 \AA\ cut off the wings of the 
$\lambda$1038 line, reproducing the peculiar line shape quite well. 
However, this analysis indicates that there should be additional flux 
visible at 1038.8 \AA; absorption from some other
atomic species may be responsible for the lack of observed emission.
The Gaussian fit to the $\lambda$1032 line is redshifted 
($v=105 \pm 11$ km s$^{-1}$, corrected for the zero-point offset of 
the wavelength calibration), with a FWHM $= 356 \pm 11$ \kms. 

The \Os\ line profile does not appear to be a standard 
type I P Cygni profile \citep{bea50}, because no blueshifted absorption 
is observed on the small but significant continuum flux around 
the emission line. While the Gaussian model fits the red wing of 
the $\lambda$1032 line quite well, there is excess emission on the 
blue wing. This asymmetry in the line-profile suggests that the 
\Os\ emission is formed in AB Aur's stellar wind, but this conclusion is
tentative until more detailed modeling is complete.

The \Ct\ $\lambda$977 line in the SiC 1b segment spectrum is shown in 
Figure~2b. Again, there are narrow \hh\ absorption lines superimposed 
on the emission feature, which were masked out before least-squares 
fitting a Gaussian to the line. The bulk of the emission is redshifted 
($v = 192 \pm 11$ \kms), but the \Ct\ emission feature clearly is not 
well-described by a single Gaussian. There is an ``absorption feature'' 
at $\sim$ 978 \AA, superimposed on the \Ct\ emission, that does not 
arise from \hh. The origin of this feature has not yet been determined, but
it is not likely to be interstellar, because of its large redshift and width.

\subsection{Molecular hydrogen}

The \hh\ absorption features visible in our \emph{FUSE} spectrum 
do not show any evidence of multiple velocity components. 
The LiF 1a and LiF 2a data were normalized with simple continua 
found from linear fits to the spectrum between 1040 \AA\ and 1120 \AA.
Our \hh\ absorption model was generated using molecular data from 
\citet{abg93a} and \citet{abg93b}. Voigt line-profiles were used
to create transmission functions, which 
were then convolved with a Gaussian instrumental line-spread function 
with $FWHM = \lambda/15000$. 
The $\chi^2$ statistic between the model 
and the data was minimized to find the best model parameters, with 
$1-\sigma$ error bars determined from the contours of $\chi^2$. 
The velocity of the gas was found by least-squares fitting Gaussians 
to a number of narrow lines between 1050 \AA\ and 1120 \AA. We measured 
$v_{\rm{H_{2}}} = 103 \pm 10$ \kms, but set this value to the 
STIS carbon monoxide velocity, $v_{\rm{CO}} = 10 \pm 3$ \kms, to 
determine the zero-point offset of our wavelength calibration. 

Since the relative populations of \hh\ energy levels with $\nu=0$ and 
$J=0,1$ are determined primarily by thermal collisions, the excitation 
temperature found from these levels should be close to the kinetic 
temperature of the gas. However, the populations of the higher levels 
($J\geq2$) can be inflated by UV and/or formation pumping and by 
radiative cascade \citep{shu82}. Therefore, the low $J$ 
and high $J$ levels were analyzed separately. Windows containing only  
lines arising from the $J=0,1,2$ levels in the five bands between 
1048 \AA\ and 1120 \AA\ were cut out 
of the normalized LiF 1a and LiF 2a data and $\chi^2$ minimization 
performed using a model containing only these lines. We permitted the
model to have a slightly different velocity shift for each of the windows, 
to eliminate any systematic errors arising from the uncertainty 
in the \hh\ velocity determination. The parameters of the model were 
$N(J=0)$, $N(J=1)$, $N(J=2)$, the column densities 
in the $J=0,1,2$ levels respectively, and $b$, the Doppler broadening 
parameter. From the ratio $N(J=1)/N(J=0)$, we calculate the gas 
kinetic temperature of the 
\hh, $T = 65 \pm 4$ K. To analyze the high $J$ lines, we again cut 
windows containing only lines arising from $J\geq2$ out of the data, 
and performed $\chi^2$ minimization using a model containing only those 
lines. We find the excitation temperature of the high $J$ lines to be 
$212 \pm 20$ K. The column densities of molecules in 
each rotational level were summed to find the total \hh\ column density,
$N(\rm{H}_2) = (6.8 \pm 0.5) \times 10^{19}$ cm$^{-2}$. The final best 
two-temperature model is shown overplotted on a portion of the LiF 1a 
spectrum in Figure~3a.

\section{STIS Observation and data reduction}

AB Aur was observed by \emph{HST} on 2000 February 28. A total exposure time
of 1813 s was obtained using the E140M echelle grating and the
$0.\arcsec2 \times 0.\arcsec06$ spectroscopic slit. 
The spectrum covered 1150--1725 \AA\ with a resolution of 
$\approx$ 46,000. The data were reduced using the STIS IDT software 
\emph{calstis}. Echelle spectra are subject to interorder scattered light; 
to compensate for this, we used the algorithm developed by Don Lindler 
and discussed in \citet{hol99}, which accurately achieves zero flux in the 
trough of the interstellar Ly$\alpha$ absorption. The STIS wavelength scale 
in this mode is accurate to 3 \kms\ \citep{pag00}. The spectrum 
between 1200 \AA\ and 1300 \AA\ is shown in Figure 1b. 
Longward of 1300 \AA, the spectrum resembles an early A star with the 
heavy line blanketing typical of Herbig Ae stars in the UV. The STIS
spectrum reached a continuum S/N=20.6 near 1471 \AA, and S/N=16 near
1504 \AA. The S/N at shorter wavelengths is lower, S/N $\leq 3.8$ per 
0.2 \AA\ bin at Lyman $\alpha$. 

\section{Analysis of STIS spectrum}

The STIS spectrum of AB Aur is rich in emission lines from a wide range of 
ionization states. The data shortward of 1310 \AA\ show type I P Cygni 
profiles arising from the low-ionization species \ion{O}{1} and \ion{N}{1}, 
and more complex emission profiles in \ion{Si}{3} and \Ct\ $\lambda$1176. 

\subsection{Extinction, Lyman $\alpha$, and $N$(\ion{H}{1})}

The measured color index of AB Aur, (B--V), is 0.12. 
IUE low-dispersion spectra of AB Aur (SWP 24389) were unreddened using 
the interstellar excinction law of \citet{car89}.
Comparison with unreddened spectral standards suggests that the 
best overall match in the FUV is $\gamma$ UMa, type A0 V (SWP 8198),
with (B--V)$_0 = 0.04$ and R = 3.1. We therefore find the selective 
extinction toward AB Aur, $\rm{E(B-V)} = 0.08$, and the general 
extinction, $\rm{A}_{V} = \rm{R} \times \rm{E(B-V)} = 0.25$.

Lyman $\alpha$ emission is formed in the chromosphere of AB Aur, and it
appears likely that the chromosphere is expanding, based on
profiles of other low-ionization species.
Since these line-profiles in the STIS spectrum of AB Aur are observed 
to be type I P Cygni profiles, with the emission extending down to 
the stellar radial velocity, it is likely that the intrinsic 
Lyman $\alpha$ profile is a type I P Cygni profile as well. 
However, the short wavelength edge of the observed emission does not 
extend down to the stellar radial velocity; this is presumably due to the 
interstellar Lyman $\alpha$ absorption. The wavelength at which the 
interstellar absorption is no longer optically thick,
where the observed flux rises above the zero level, 
is highly sensitive to the interstellar \ion{H}{1} column density,
but insensitive to the assumed b-value. We find $N($\ion{H}{1}$)$ toward 
the star to be $3 \times 10^{20}$ cm$^{-2}$. We obtain roughly the
same value using several different assumptions about the underlying
continuum flux. However, uncertainty about the exact shape of the 
stellar line-profile could cause the error in this value to be as 
much as 50\%. Future work will refine $N($\ion{H}{1}$)$ after detailed
modeling of AB Aur's winds determines the unattenuated stellar flux 
presented to the interstellar \ion{H}{1} atoms. 

\subsection{Carbon monoxide}

The (1-0), (2-0), and (3-0) bands of the CO Fourth Positive band system 
($A \: ^{1}\Pi \:-\: X \:^{1}\Sigma ^{+}$) are
visible in the STIS spectrum (Figure~3b). No lines arising from 
$^{13}$CO were observed. The continua in the vicinity of the bands were 
fitted with fifth-degree polynomials. An absorption model was generated 
using wavelengths and oscillator strengths from \citet{mor94} and 
energies of the ground-state levels were calculated using the Dunham 
coefficients from \citet{far91}. $\chi^2$ minimization was performed on all 
three bands simultaneously to find the best rotational excitation temperature, 
$T_{\rm{CO}} = 7.0 \pm 0.7$ K, Doppler broadening parameter, 
$b = 3.8 \pm 0.4$ \kms, column density of $^{12}$CO, 
$N(\rm{CO}) = (7.1 \pm 0.5) \times 10^{13} \ \rm{cm}^{-2}$, 
and velocity centroid, $v = 10 \pm 3$ \kms.

\section{Discussion}

We find a total column density of hydrogen atoms, $N_{\rm{H}} = 
N($\ion{H}{1}$) + 2N(\rm{H}_2) = 4.4 \times 10^{20}$ cm$^{-2}$, 
which is typical of a diffuse molecular cloud. The fractional abundance 
of \hh, $f = 2N(\rm{H}_2) / N_{\rm{H}} = 0.31$. Only 11
out of 91 stars in the \citet{sav77} \emph{Copernicus} survey of 
interstellar \hh\ show a larger value of $f$, and these 
sight-lines all have larger values of $\rm{E(B-V)}$ and $N_{\rm{H}}$.
This might suggest that our line-of-sight to AB Aur has an 
anomalously high molecular fraction, but this conclusion is tentative 
until further \emph{FUSE} \hh\ measurements of a variety of interstellar
environments are complete. The CO$/$\hh\ ratio is $(1.04 \pm 0.11) 
\times 10^{-6}$. We find the gas-to-dust ratio, $N_{\rm{H}}/\rm{E(B-V)} = 
5.5 \times 10^{21}$. The mean ratio for standard clouds within 2 kpc of the 
Sun is $5.9 \times 10^{21}$, indicating that our gas-to-dust 
ratio is not atypical \citep{boh78}. 

Optical coronagraphic images of AB Aur show an extended arc of 
reflection nebulosity to the east of the system, in addition to the 
circumstellar disk centered on the star \citep{nak95, gra99}. 
\citet{nak95} suggest that such nebulosities are the
remnants of the original molecular clouds that collapsed to form
low-mass stars. Our values of the various parameters describing the cold \hh\
and CO gasses ($T_{\rm{kin}}$, $N_{\rm{H}}$, $f$, CO$/$\hh\, and 
$N_{\rm{H}}/\rm{E(B-V)}$) are consistent with this suggestion, and we
believe that the \hh\ and CO gasses toward AB Aur are probably located in a 
remnant molecular cloud envelope around the star. The fact that the 
velocity of the CO gas is equal (within measurement error) to the 
stellar radial velocity supports this conclusion. The molecular gasses
are not likely to be located in the circumstellar disk around AB Aur, since 
the disk inclination is such that our line-of-sight to the star
does not pass through the disk \citep{gra99}.

Turning to the hot, highly-ionized gasses, FUV emission lines from 
\Os\ and \Ct\ are observed in the solar spectrum, and are produced
in the transition zone between the chromosphere and the corona. However,
detailed modeling of the \ion{N}{5} ($T \approx 1.4 \times 10^5$ K) 
emission previously observed
from AB Aur with GHRS found that a homogeneous, spherically-symmetric 
high-temperature zone could not produce the observed \ion{N}{5} 
emission without producing strong \ion{N}{4} or \ion{C}{4} emission
as well, which is not seen \citep{bou97}. \citet{bou97} concluded that
small, hot clumps, formed by shocks in a stellar wind, could reproduce 
the FUV emission lines, as well as the weak X-ray flux observed from AB Aur.
Future work will determine if the wind-shock model can reproduce the 
observed emission from \Ct\ and the higher-temperature 
($T \approx 3 \times 10^5$ K) species \Os.

This work is based on observations made with the NASA-CNES-CSA 
\emph{Far Ultraviolet Spectroscopic Explorer} and on 
NASA-ESA \emph{Hubble Space Telescope} observations obtained at the 
STScI under the Guaranteed Time Observer program 8065. \emph{FUSE} is 
operated for NASA by the Johns Hopkins University under NASA 
contract NAS5-32985. The work performed by J.-C. Bouret was 
supported by a National Research Council-(NASA GSFC) Research Associateship.

\clearpage

\begin{figure}[ht]
\begin{center}
\epsfig{file=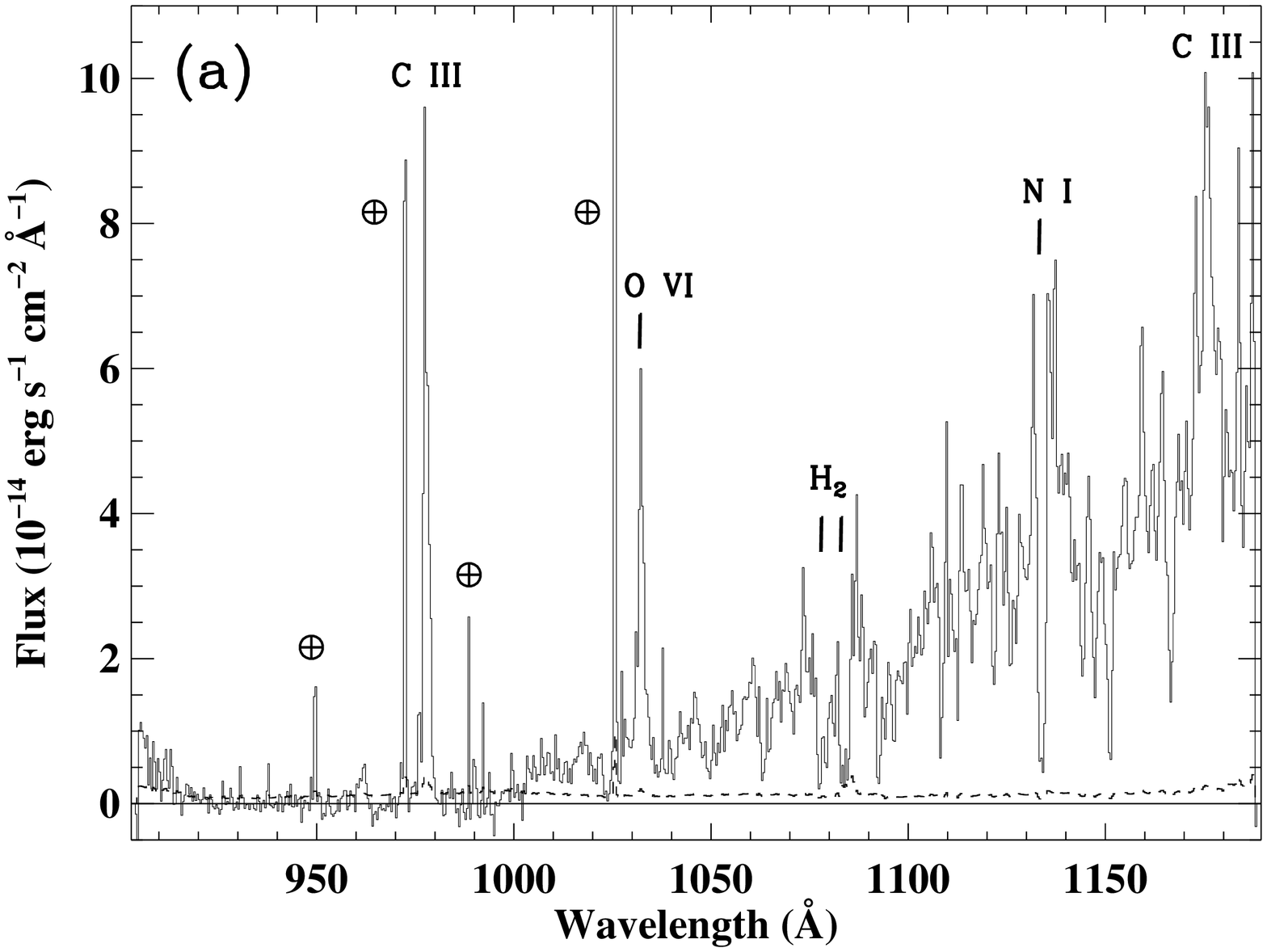, height=3.6in, angle=90}
\epsfig{file=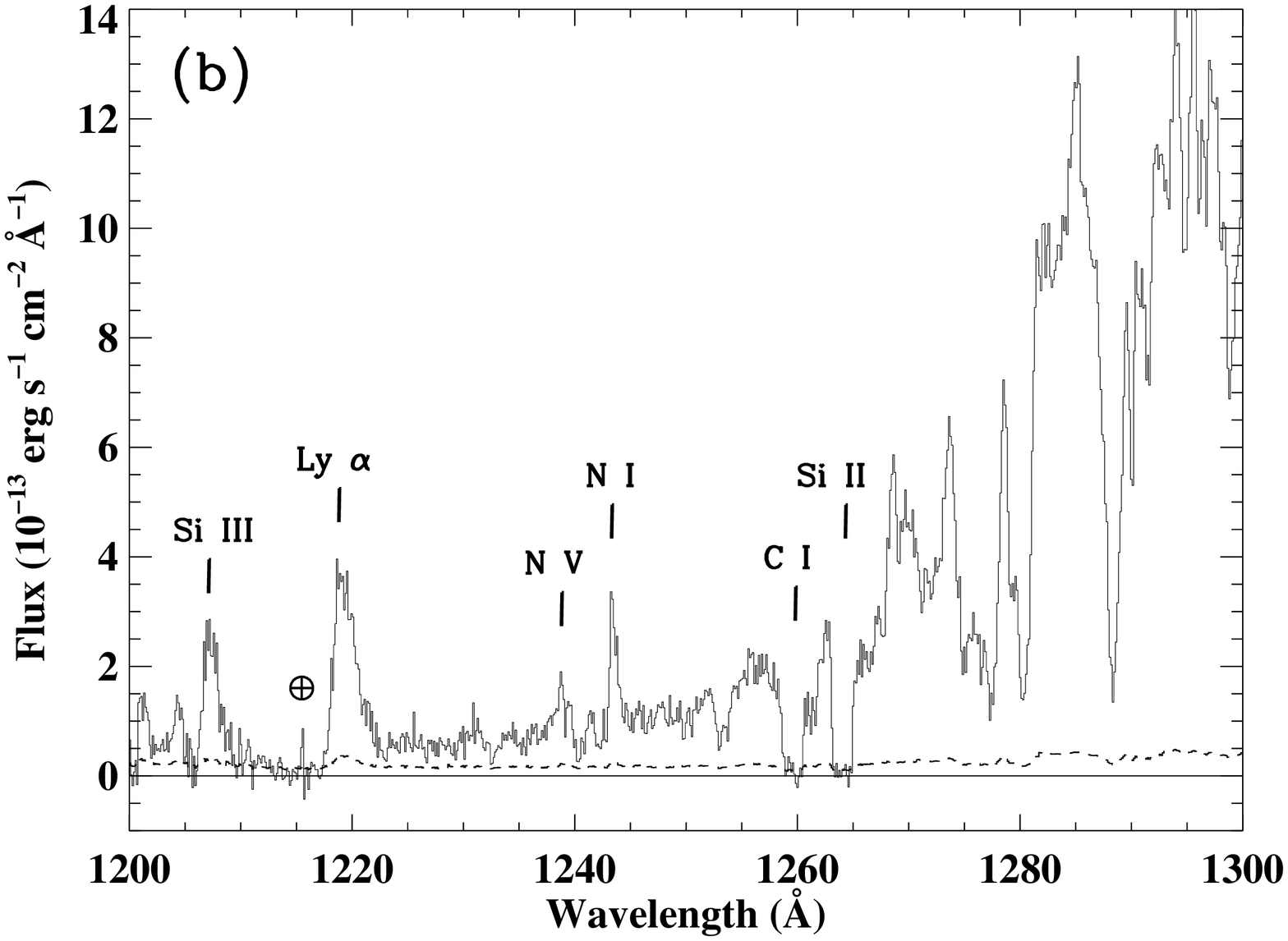, height=3.6in, angle=90}
\caption{Overview of the AB Aurigae FUV spectrum. The $1-\sigma$ flux 
measurement errors are plotted with a dashed line. Airglow lines are 
indicated by $\oplus$, and prominent features are labeled. \ \ \textbf{(a)} 
\ \emph{FUSE} spectrum of AB Aur, combining data from all channels. 
The data have been rebinned by a factor of 80 for this plot.  \ \ \textbf{(b)} 
\ Short wavelength portion of the STIS E140M spectrum of AB Aur. The data have 
been rebinned by a factor of 10 for this plot.}
\end{center}
\end{figure}

\noindent
\vskip -0.4in
\begin{figure}[ht]
\begin{center}
\epsfig{file=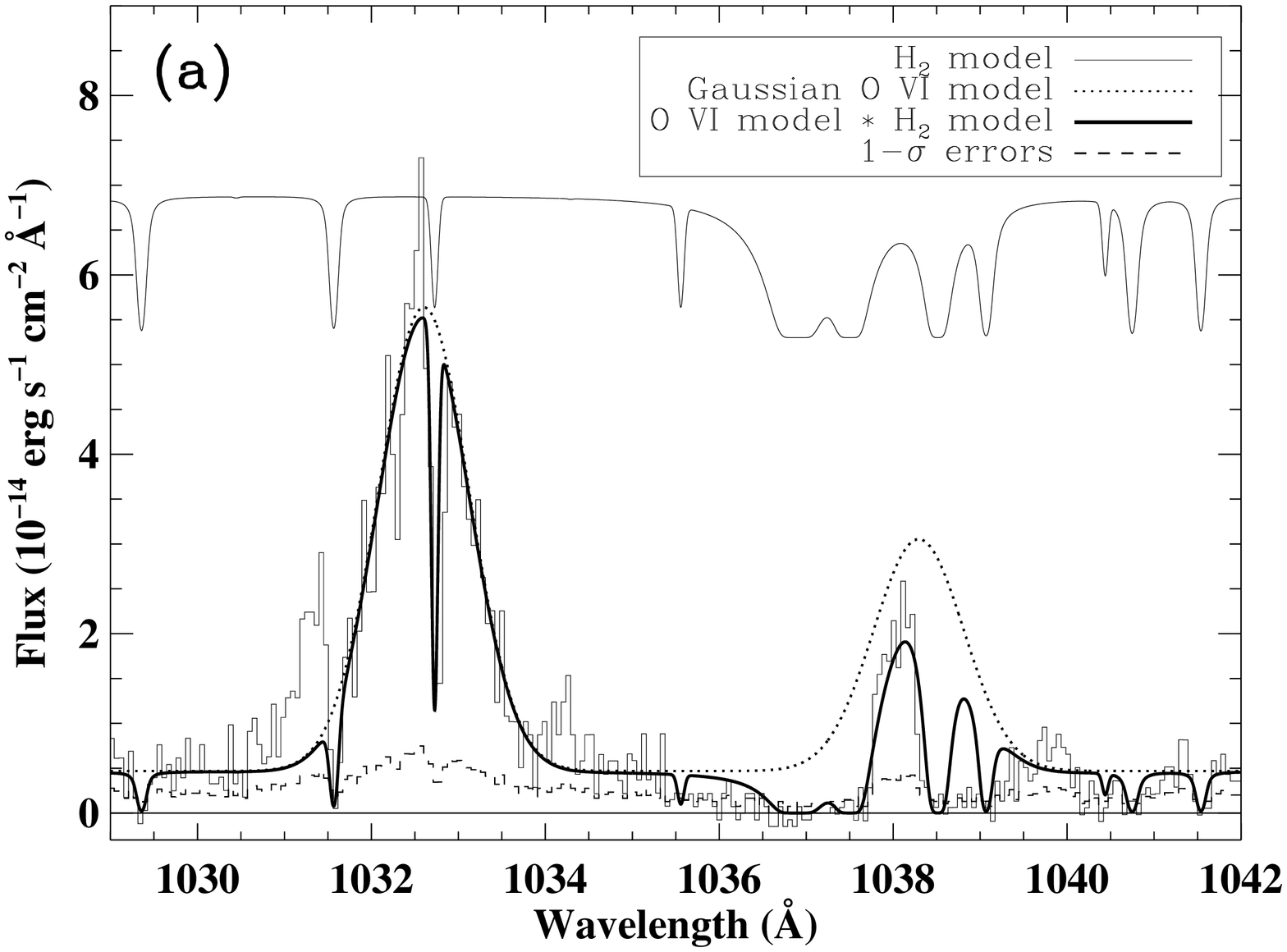, height=3.6in, angle=90}
\epsfig{file=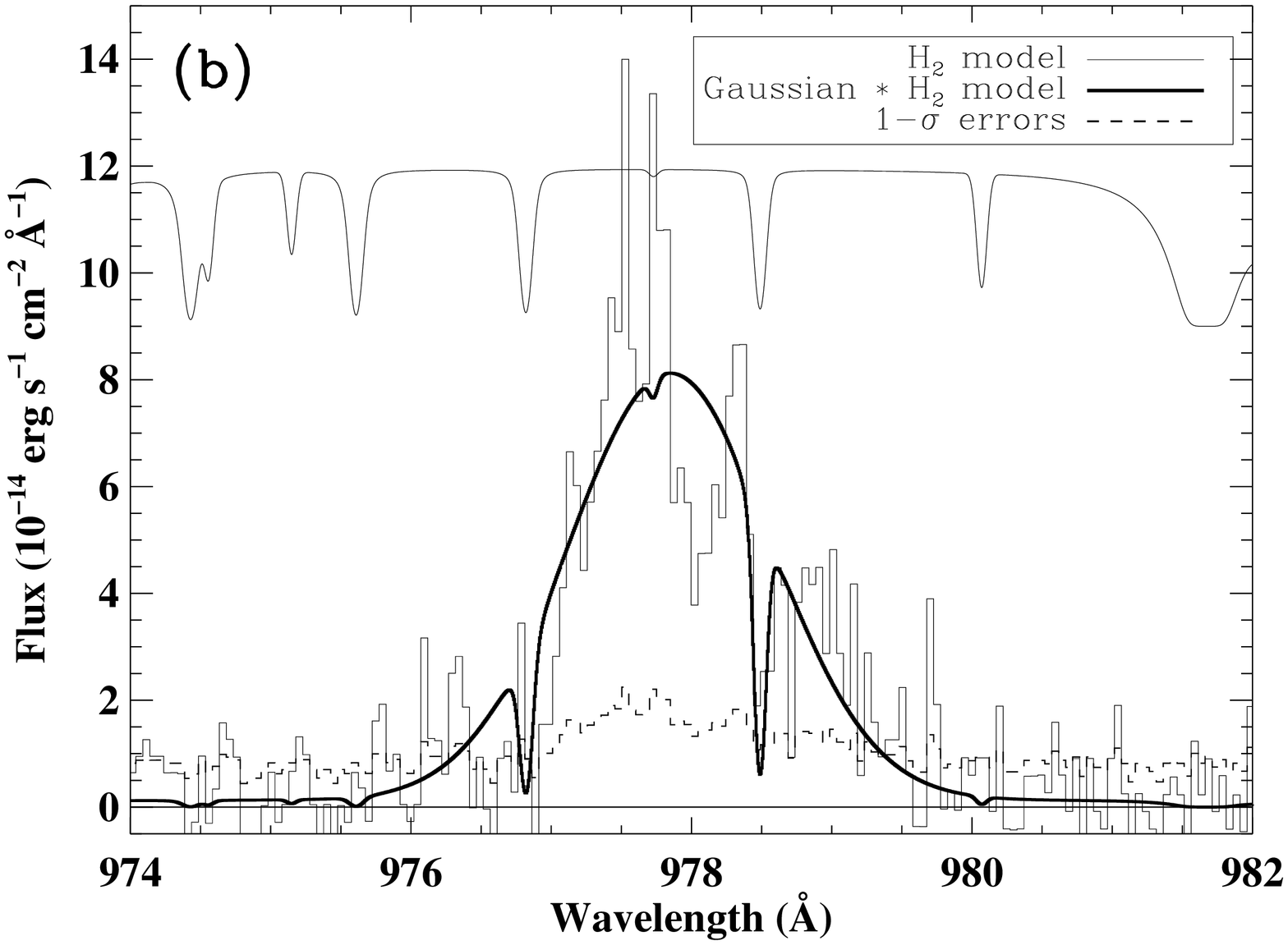, height=3.6in, angle=90}
\caption{Emission from highly-ionized species in the \emph{FUSE} 
spectrum. The data have been rebinned by a factor of 8 for these
plots. Our best \hh\ absorption model is shown at the top of both plots.
The $1-\sigma$ flux measurement errors are plotted with a dashed line.
\ \ \textbf{(a)} \ \Os\ $\lambda \lambda 1032, 1038$ doublet in the 
LiF 1a spectrum. An optically thin Gaussian \Os\ model is shown with a 
dotted line. The normalized \hh\ model was multiplied by the \Os\ 
model and overplotted with a thick line. \ \ \textbf{(b)} \ \Ct\ $\lambda$977
line in the SiC 1b spectrum. Our Gaussian fit to the line multiplied by 
the \hh\ model is overplotted with a thick line.}
\end{center}
\end{figure}

\noindent
\begin{figure}[ht]
\begin{center}
\epsfig{file=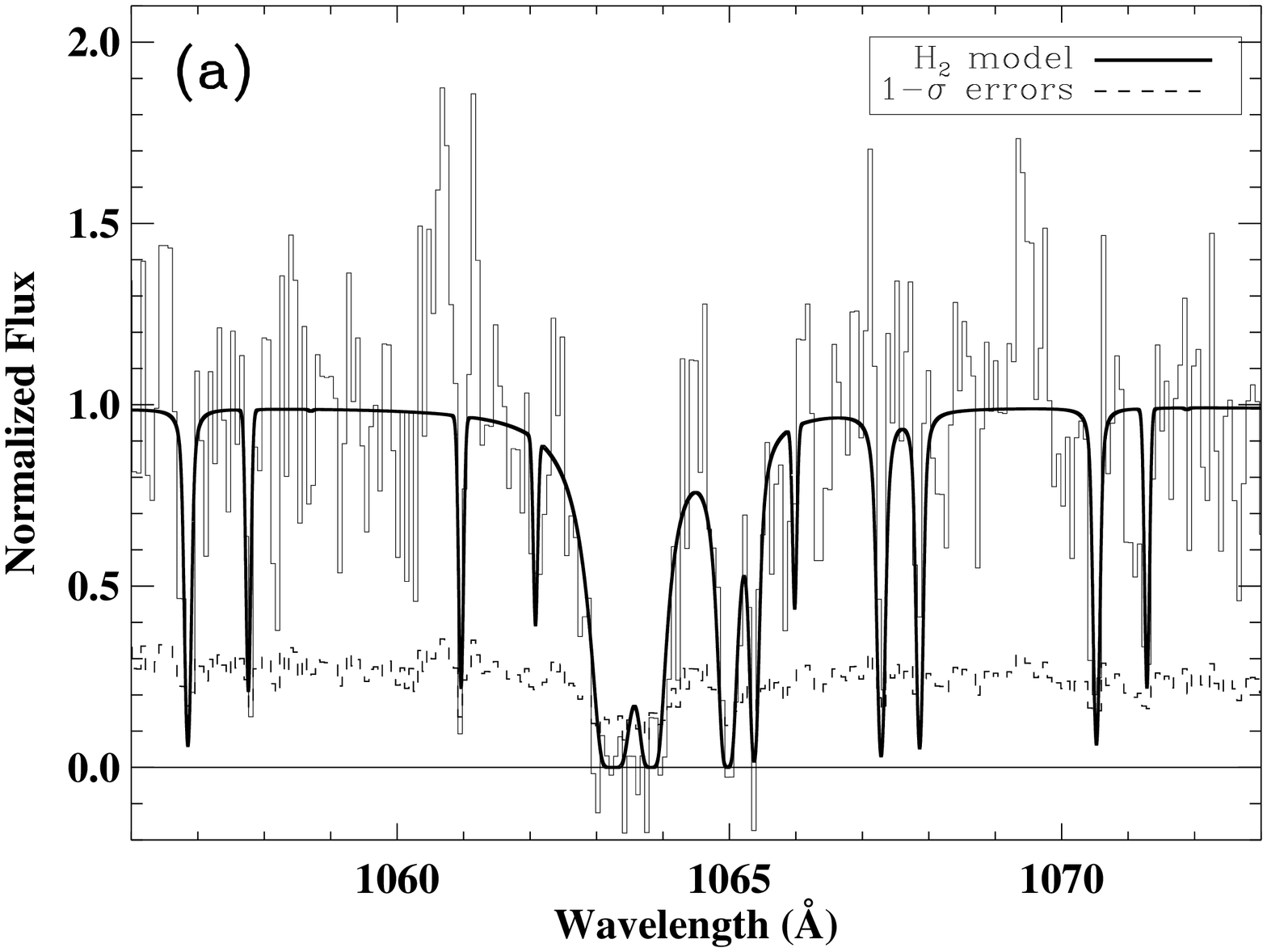, height=3.65in, angle=90}
\epsfig{file=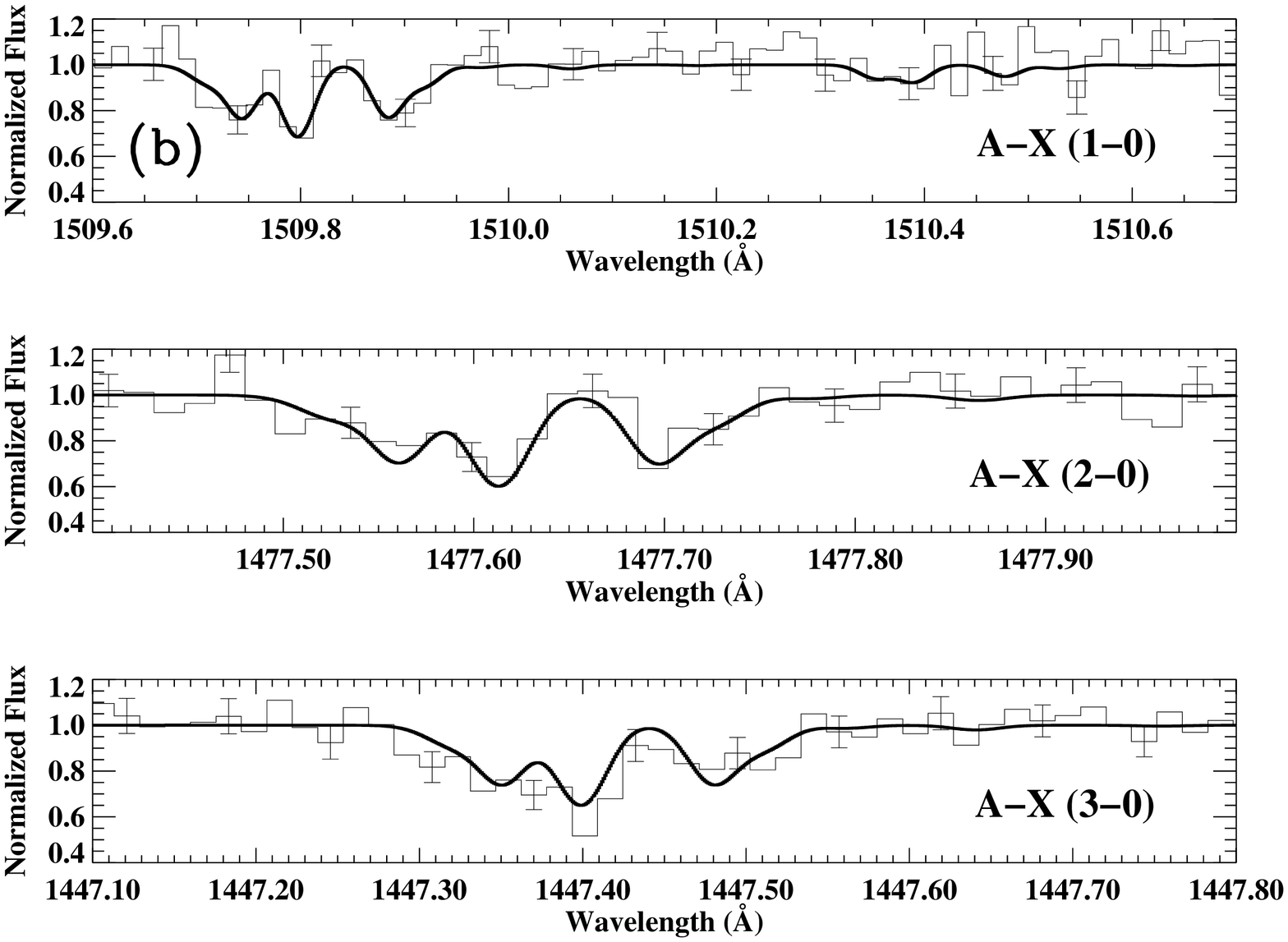, height=3.55in, width=2.59in, angle=90}
\caption{Molecular gasses toward AB Aur. \ \ \textbf{(a)}  \ \hh\ Lyman
(3-0) band in \emph{FUSE} LiF 1a spectrum. The data have been rebinned by
a factor of 10 for this plot. The $1-\sigma$ flux measurement errors are 
plotted with a dashed line. The best two-temperature model is
overplotted with a thick line. \ \ \textbf{(b)} \ Three bands of the CO 
Fourth Positive band system in the STIS E140M spectrum. The data have not 
been rebinned or smoothed for this plot. The $2-\sigma$ flux error bars 
are overlaid on the data. The best model is overplotted with a thick line.}
\end{center}
\end{figure}

\end{document}